\documentclass{article}
\usepackage{graphicx}% Include figure files
\usepackage{dcolumn}% Align table columns on decimal point
\usepackage{bm}% bold math
\usepackage{graphics}
\begin{document}
\title{Scaling at the selective withdrawal transition}
\author{Itai Cohen, Sidney R. Nagel}
%\email{icohen@deas.harvard.edu}
%\affiliation{James Franck Institute, University of Chicago, Chicago, IL 60637}
\begin{abstract}
In the selective withdrawal experiment fluid is withdrawn through a tube with its tip 
suspended a distance $S$ above an unperturbed two-fluid interface.  
At low withdrawal rates, 
$Q$, the interface forms a steady state hump and only the upper fluid is withdrawn.  When 
$Q$ is increased (or $S$ decreased), the interface undergoes a transition so that the lower fluid 
is entrained with the upper one, forming a thin steady-state spout.  Near this 
discontinuous transition the hump curvature becomes very large and displays power-law 
scaling behavior.  This scaling is used to show that steady-state profiles for 
humps at different flow rates and tube heights can all be scaled onto a single similarity 
profile. 
\end{abstract}
\maketitle
Is it possible to classify topological transitions in nonlinear 
fluid systems \cite{Ca93,Bertozzi94,Goldstein93,Pugh98} in 
the same manner as one classifies thermodynamic transitions?  When the topological 
transition involves formation of a singularity in the fluid flows or interface shapes, a 
similarity solution can provide a simplified description of the flows and help make 
such a classification \cite{Barenblatt96}.   A crucial component to this approach 
involves determining how 
characteristic physical quantities and lengths describing the fluid system scale near the 
singularity.  In many cases these singularities manifest themselves in the transition 
dynamics \cite{JRL&HAS98,Bensimon86,Sid&Lene00,Lathrope00} and do not appear 
in the steady state flows.  Here, we report on
steady-state interface profiles near the topological transition associated with the selective 
withdrawal experiment.  Despite the transition being discontinuous, scaling of the 
interface is observed as the transition is approached.  

\begin{figure}[htbp]
\centerline{
\resizebox{0.75\textwidth}{!}
{\includegraphics{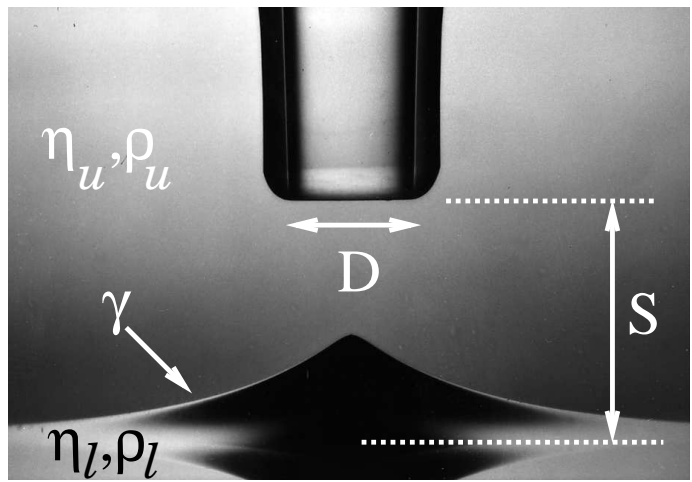}}}
\caption[]{Photograph of two-fluid interface forming a hump.  The parameters for this 
problem include the height of the tube from the unperturbed interface, $S$, withdrawal rate, $Q$,  
surface tension of the two-fluid interface, $\gamma$, fluid densities, $\rho_{u}$ 
and  $\rho_{l}$, fluid 
viscosities, $\eta_{u}$ and $\eta_{l}$, tube diameter, $D$, fluid 
height above the interface, fluid height below the interface, container size, and surfactant 
concentration.  Here we restrict the investigation to the $S$ vs. $Q$ parameter 
space and defer discussion of the remaining parameters to \cite{tobe}.}
\label{fig:SW_parameters}
\end{figure}

In the selective withdrawal experiment a tube is immersed in a filled container so 
that its tip is suspended a height $S$ above an unperturbed interface separating two 
immiscible fluids.  
When fluid is pumped out through the tube at low flow rates, $Q$, only 
the upper fluid is withdrawn and the interface is deformed into 
an axi-symmetric steady-state hump (Fig. \ref{fig:SW_parameters}) due to 
the flows in the upper fluid.  The hump grows 
in height and curvature as $Q$ increases or $S$ decreases until the flows undergo a transition 
where the lower fluid becomes entrained in a thin axi-symmetric spout along with the 
upper fluid.  The two-fluid interface becomes unbounded in the vertical direction 
thus changing the 
topology of the steady state.  Once the spout has formed, an increase in $Q$ or decrease in 
$S$ thickens the spout.  

The interfacial profiles at different flow rates and tube heights 
are recorded.  Near the transition, the steady-state radius of 
curvature of the hump tip is orders of magnitude smaller than the length 
scales characterizing the boundary conditions (e.g. the tube diameter, $D$).  
This separation of length scales suggests that a similarity analysis of 
the steady-state hump profiles might be possible. However, for the range of 
parameters explored thus far, even when the system is arbitrarily close 
to the transition from hump to spout, 
the mean curvature of the hump tip, $\kappa$, while large, remains finite. 
Nevertheless, by fixing $S$ and looking at the steady-state profiles as $Q$ is 
increased, we observe that both the hump curvature and height display scaling behavior 
characteristic of systems approaching a singularity.  Since the divergence 
is cut off before a singularity is reached this transition appears 
to be ``weakly-first-order.''

As shown in Fig. \ref{fig:SW_parameters}, the parameters important for this experiment are the upper 
and lower fluid viscosities and densities ($\eta_{u}, \eta_{l}, \rho_{u}, \rho_{l}$), interfacial tension ($\gamma$), orifice diameter ($D$), tube height ($S$), and flow rate ($Q$).  In looking 
for scaling of the steady-state profiles, care must be taken to design an experimental 
apparatus capable of isolating the profiles near the transition.  Experiments were 
performed in large tanks (30cm $\times$ 30cm $\times$ 30cm) capable of holding 
fluid layers that were 
each about 12 cm in height.  To ensure that the upper fluid level remained 
constant, the withdrawn fluid was recycled back into the container (the bottom 
fluid layer thickness remains constant when the system is in the hump state).  Steady 
withdrawal rates were achieved by using a gear pump.  We verified that for the tube 
diameter ($D = 0.16$ cm), tube heights ($0.1$ cm $\leq S \leq 1.1$ cm), and flow 
rates ($Q \leq 10$ ml/sec) used in the experiments, the container walls 
were sufficiently distant and the fluid 
layers sufficiently thick so as not to affect the flows \cite{tobe}.  We 
measured the upper(Heavy Mineral Oil) and lower(Glycerin-Water mixture) 
fluid viscosities to be 
$\eta_{u} = 2.29$ St and $\eta_{l} = 1.90$ St, the upper and lower fluid 
densities to be $\rho_{u} = 0.88$ g/ml and $\rho_{l} = 1.24$ g/ml, and the 
surface tension\cite{Neumann,Hansen} to be $\gamma = 31$ dynes/cm.
Attempts to increase $\kappa$ near the transition by decreasing 
$\gamma$ cause fluid mixing and result in a diffuse interface 
at high shear rates so that
some fraction of the lower fluid is always being withdrawn \cite{JRL89}.    

While many of the parameters mentioned influence the flows, our 
understanding of the scaling behavior can be conveyed by focusing on $S$ 
and $Q$.  We can fix $Q$ and track the development of the hump 
profiles as a function of $S$.  Below the tube height, $S_{u}$, the hump is unstable and 
undergoes a transition to a spout.  Figure \ref{fig:Su,Ku,hu,DeltaSvsQ} shows that $S_{u} \propto Q^{0.30 \pm 0.05}$ 
\cite{Svs.Q}.  At low $Q$ the transition is hysteretic: the 
value of $S$ where the spout becomes unstable and decays back into the hump is
larger than $S_{u}$.  We define the difference of the two heights or hysteresis as $\Delta S$.  
Figure \ref{fig:Su,Ku,hu,DeltaSvsQ} indicates that the data is consistent with an exponential decrease: $\Delta S =  0.04 \exp^{-
Q/0.032}$.  For $Q > 0.1$ ml/sec, $\Delta S$ was too small to measure. 
\begin{figure}[htbp]
\centerline{
\resizebox{0.75\textwidth}{!}
{\includegraphics{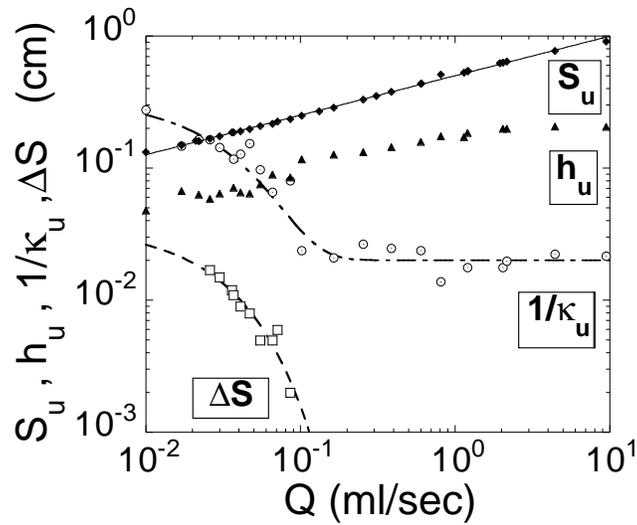}}}
\caption[]{Plots of the transition tube height, $S_{u}$, hysteresis, $\Delta S$, 
hump height, $h_{u}$, and mean radius 
of curvature $1/\kappa_{u}$, as a function of $Q$.  
We find that $S_{u} \propto Q^{0.30 \pm 0.05}$ 
(solid line).  The $\Delta S$ data is fit with an exponential decay 
(dashed) of the form: $\Delta S = 
0.04 \exp^{-Q/0.032}$ although the range is not sufficient to exclude a power-law decay.  
We fit $1/\kappa_{u}$ (dash dot), with the form: $1/\kappa_{u} = 0.02 + 0.32 \exp^{-Q/0.032}$.}
\label{fig:Su,Ku,hu,DeltaSvsQ}
\end{figure}

In order to measure the mean curvature of the hump tip, $\kappa$, we first fit the tip of the 
recorded profile with a Gaussian and then calculate the curvature of the fitting function at 
the hump tip.  Figure \ref{fig:Su,Ku,hu,DeltaSvsQ} also shows the hump height, $h_{u}$, and mean radius of 
curvature, $1/\kappa_{u}$, at 
the transition as a function of $Q$.  The dramatic decrease in $\Delta S$ coincides with the 
onset of a flat asymptotic dependence for $1/\kappa_{u}$ at $Q > 0.1$ ml/sec.  
We quantify this correlation, by fitting the curvature data with 
the form $1/\kappa_{u} = 0.02 + 0.32 \exp^{-Q/0.032}$ which has the same exponential decay with $Q$ 
as does $\Delta S$.  For $Q > 0.1$, ml/sec we find both $\kappa_{u}$, and $h_{u}$ to be 
independent of the orifice diameter $D$ \cite{tobe}. We restrict our scaling analysis to this 
regime.\cite{Surfac_disc}

Figure 3a plots $\kappa$ (where $0 \leq \kappa \leq \kappa_{u}$), 
versus $Q$ for six data sets corresponding to different 
values of $S$.  As shown in the inset of Fig. 3a, all fifteen data 
sets display a power-law divergence for $\kappa$, as $Q$ 
approaches $Q_{c}$ (a fitting parameter). While the power-law 
exponents remain constant as $S$ is varied, the prefactors to 
the power laws, $c_{\kappa}(S)$, vary slightly with $S$ 
and are scaled out in the inset. Figure 3b plots the hump 
height, $h_{max}$, versus $Q$.  The inset to Fig. 3b shows that as $Q$ 
approaches $Q_{c}$ (obtained from Fig. 3a inset), the hump height 
approaches $h_{c}$ (a fitting parameter) as a power law.  Once again, 
the power-law exponents for these data sets are the same over this range of $S$.  
The prefactors, $c_{h}(S)$, are scaled out in the inset. Note that $Q_{c}$ changes 
with $S$ indicating that the system can approach a continuous line of divergences. 
Combining the two scaling dependencies in Fig. 3c, we
plot $(h_{c}-h_{max})/h_{max}$ versus the normalized curvature, $\kappa/n$.  
We find that $(h_{c}-h_{max})/h_{max}$ 
scales as $(\kappa/n)^{0.85 \pm 0.09}$ indicating that even though both $h_{c}$ 
and the power-law prefactor, $n$, change with $S$, the power-law exponents 
are independent of $S$ for 
this range of tube heights. Note that $n(S) = c_{h}(S) [c_{\kappa}(S)^{0.85}]$. 
The transition cuts off the evolution of the hump states making it impossible for the 
system to approach arbitrarily close to the singularity and limiting 
the precision with which we can determine the exponents.
\begin{figure}[htbp]
\resizebox{0.5\textwidth}{!}
{\includegraphics{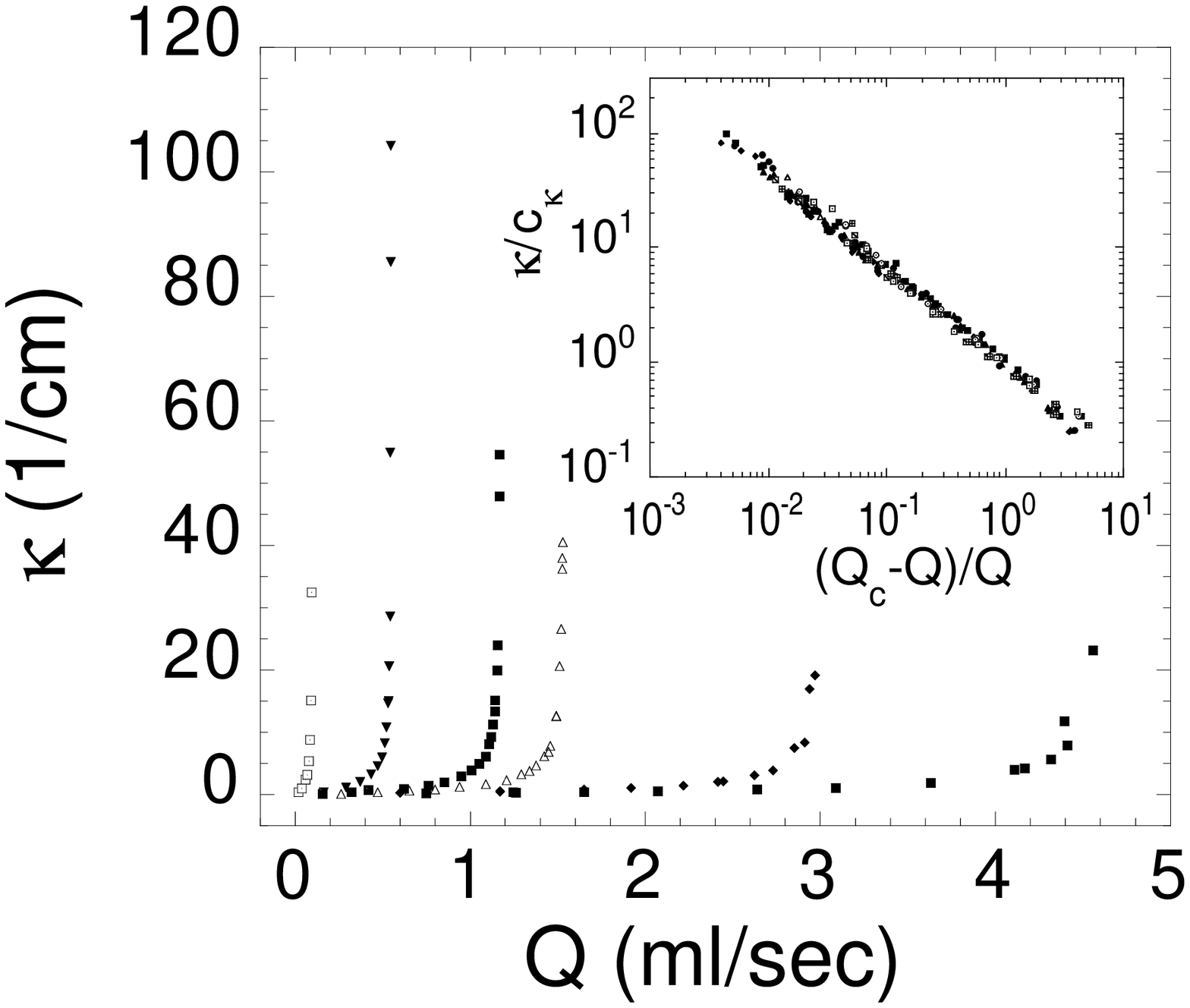}}
\vspace{4mm}
\resizebox{0.5\textwidth}{!}
{\includegraphics{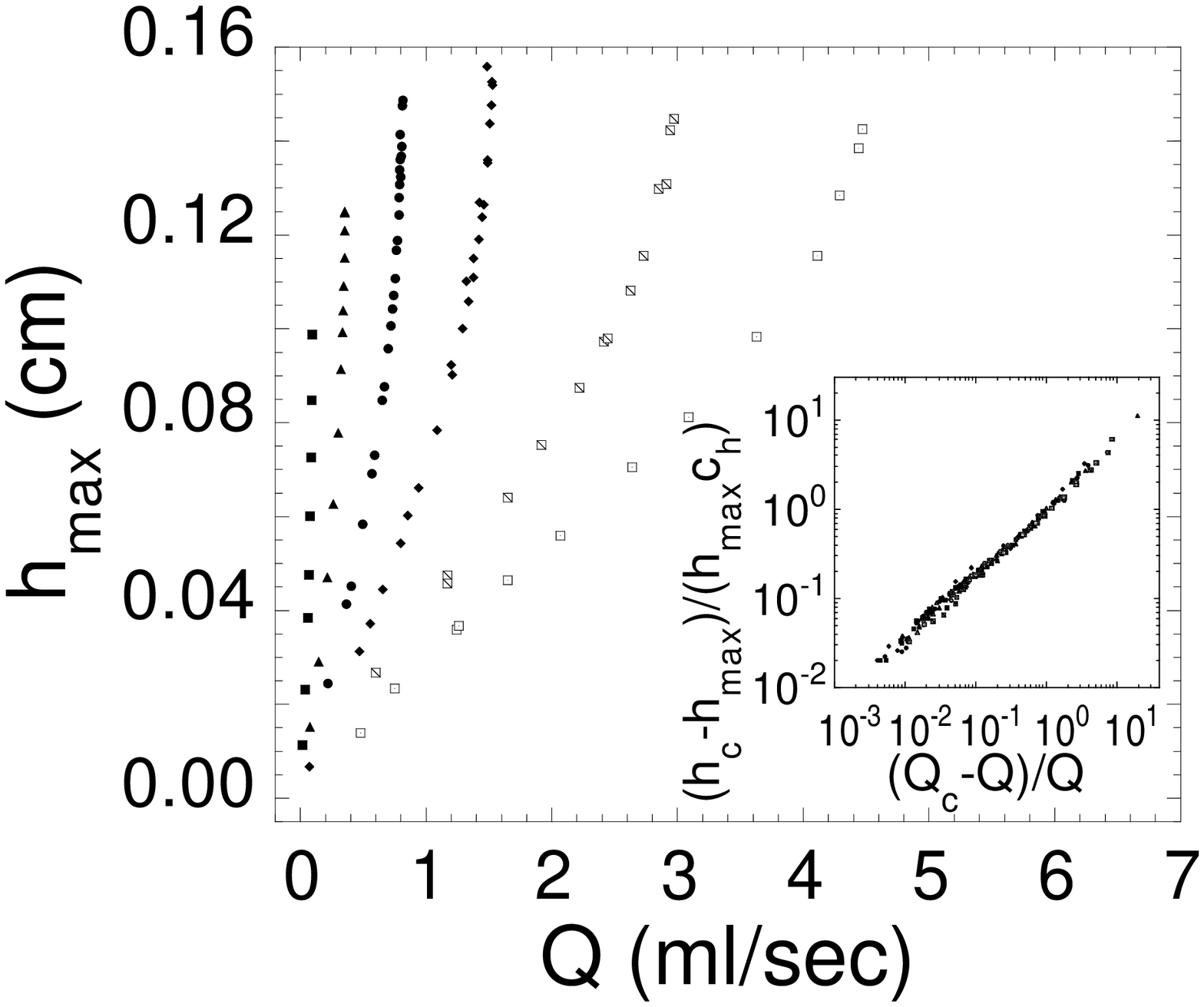}}
\vspace{4mm}
\resizebox{0.5\textwidth}{!}
{\includegraphics{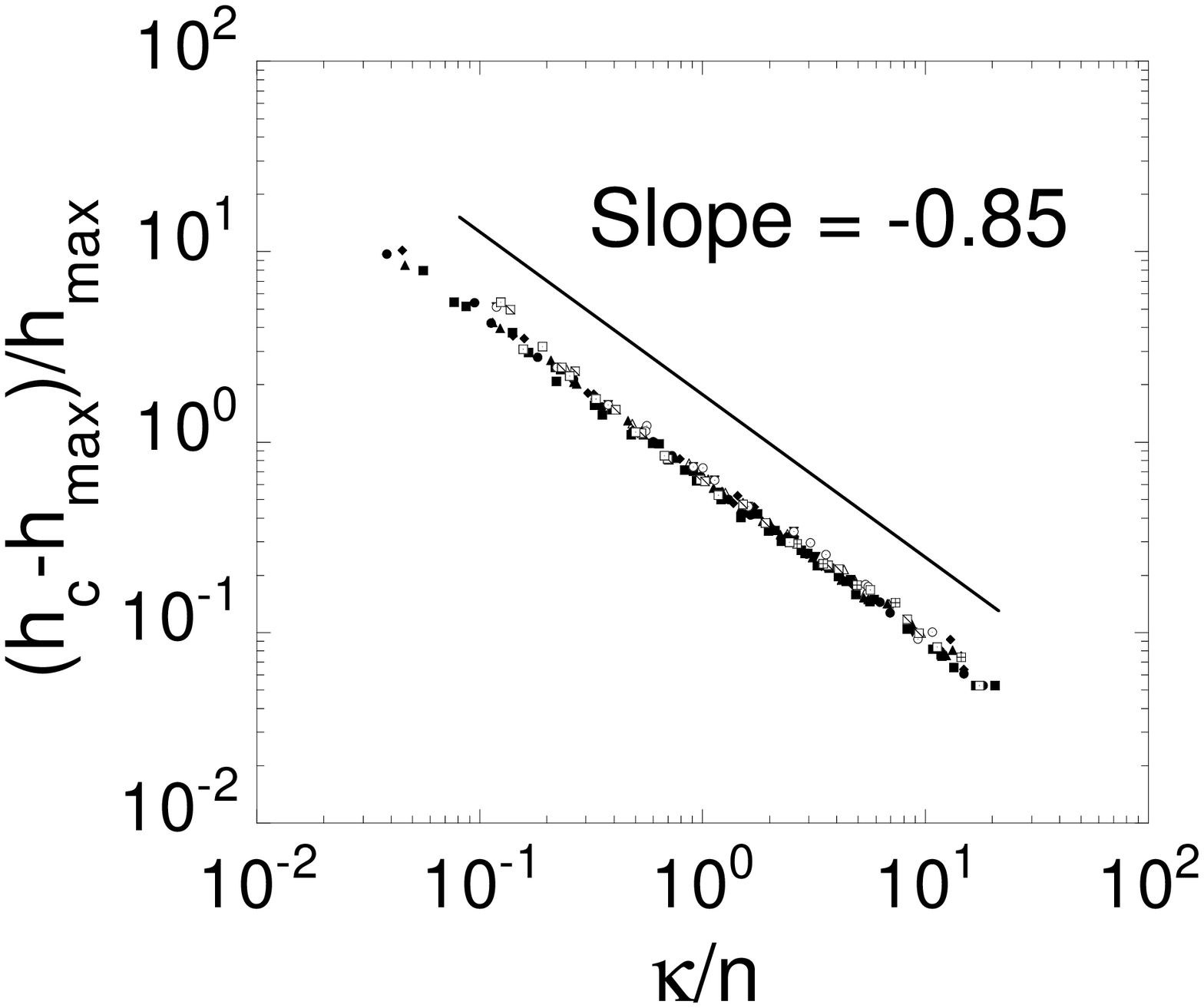}}
\caption[]{
Scaling for the mean hump tip curvature, $\kappa$, and height $h_{max}$.  3a 
plots $\kappa$ vs. $Q$ for six tube heights.  Each curve displays 
diverging behavior with increasing $Q$.  For each tube height we choose a critical flow rate 
$Q_{c}$ as a fitting parameter and show (3a inset) that as $Q$ approaches $Q_{c}$ the curvatures 
increase as power laws.  The prefactors to the curvature power laws, $c_{\kappa}(S)$, are scaled out  
in the inset.  Figure 3b plots $h_{max}$ vs. $Q$ for six tube heights.  For each $S$ 
we choose a critical hump height $h_{c}$ as a fitting parameter. As $Q$ 
approaches $Q_{c}$ (taken from 3a inset), the hump heights approach the critical heights as 
power-laws.  The prefactors, $c_{h}(S)$, to the power laws in the inset are scaled out.  Figure 
3c plots $(h_{c}-h_{max})/h_{max}$ vs. $\kappa/n$ for the entire data set corresponding to fifteen 
different straw heights.  The prefactor $n$ 
roughly decreases as $\exp^{-2.5 S}$.  The line corresponds to a 
power-law with an exponent of -0.85.}
\label{fig:Scaling}
\end{figure}

The scaling observed for $h_{max}$ and $\kappa$ suggests that the hump 
profiles should display universal behavior as $h_{max}$ nears $h_{c}$.  The quantities $1/(\kappa/n)$ 
and $(h_{c}-h_{max})/h_{max}$ track how quickly the radial and axial length scales decrease as the 
system approaches the singularity and are therefore used to scale the profiles.  We define 
the scaled variables:
\begin{equation}
\label{similarity_var}
H(R) = \frac{h_{c}-h(r)}{h_{c}-h_{max}} \qquad and \qquad R = \frac{r\kappa}{n},
\end{equation}
where $h(r)$ is the hump profile and $h_c$ is taken from Fig. 3.  
The transformation shifts the profiles so that under 
scaling the singularity occurs at the origin and the maximum hump heights occurs at $H = 1$ 
and $R = 0$.  Figure 4 shows eight scaled profiles for the $S = 0.830$ cm data set. 
In the bottom inset we overlay the hump profiles scaled in the main figure.  
We find excellent collapse of the profiles. The solid line in Fig. 4 is a power law that fits 
the data in the region beyond the parabolic tip.  The picture that emerges is of a parabolic 
tip region which decreases in its radial scale and is simultaneously pulled towards the 
singularity in the axial direction leaving behind it a power-law profile with an exponent of 
$0.72 \pm 0.08$.  This exponent is within error (although slightly smaller) of the exponent 
observed in the scaling relation of Fig. 3c which predicts an exponent of $0.85 \pm 0.09$.

Typically, the observed scaling dependencies in these types of problems result from
the local stress balance.  A scaling analysis where the viscous stresses of the upper and 
lower fluids balance the stress arising from the interfacial curvature predicts 
linear scaling dependencies and conical profile shapes.  The non-linearity of the observed 
dependencies indicates that either a different stress balance 
governs the flows (e.g. only viscous stress due to upper fluid balances 
stress due to the interface curvature) or that non-local effects 
are coupling
into the solution.  A more detailed discussion can be found in \cite{tobe}.

Finally, we compare the similarity curves for five different tube heights in the upper right 
inset of Fig. 4.  The profiles all display the 
same power-law dependence.  Within error, the normalized curvature $\kappa/n$ (taken from 
Fig. 3c) can be used to scale the radial components of these profiles and obtain good collapse.  
In Fig. 3c we find that the normalization prefactors, $n$, decrease as $\exp^{-2.5 S}$.  Here, we 
correlate this decrease with the observation that the profiles become shallower at larger $S$.  
The points of deviation for the $S = 0.255$ cm and $0.381$ cm profiles mark the transition 
from the similarity regime to the matching regime beyond which the profiles become 
asymptotically flat.  At large enough radii all of the scaled profiles display these 
deviations.

We have shown that in the $Q > 0.1$, ml/sec regime, a 
similarity analysis can be used to describe the flows near the selective-withdrawal 
transition \cite{Acrivose78}.  We have observed power-law 
scaling of the hump height and 
curvature (Fig. 3) and used these scaling relations to collapse 
the hump profiles at different flow rates and tube heights onto a single universal curve 
(Fig. 4).   However, the origin of the saturation of $\kappa$ at large $Q$ remains an 
important unexplained problem.
Further insight into this cutoff behavior may be gained by comparing with an 
analogous two-dimensional (2-D) problem which roughly corresponds to replacing the 
tube with a line sink.  Jeong and Moffatt \cite{Moffatt92} showed that in an 
idealized case where the 
bottom fluid is inviscid while the top fluid is very viscous, the 2-D hump interface forms 
a 2-D cusp singularity above sufficiently high withdrawal rates.  Recently, 
Eggers \cite{Eggers01} showed that the solution changes when the lower fluid has a finite viscosity; 
the system no longer manifests a singularity \cite{EggersRev97}.  Instead, the 
approach to the singularity 
is cut off and the system undergoes a transition to a different steady state.  In this new 
state, a sheet of the lower fluid is entrained along with the upper fluid into the line sink.  
However, the finite lower fluid viscosity prevents the hump profiles from scaling onto a 
similarity solution.  
\begin{figure}[tp]
\centerline{
%{\epsfig{file=/home/icohen/icohen/Selective_Withdrawal/PRL_paper/Figures/Fig.4_09.11.01.eps,width=0.4}}}
\resizebox{0.75\textwidth}{!}
{\includegraphics[clip=]{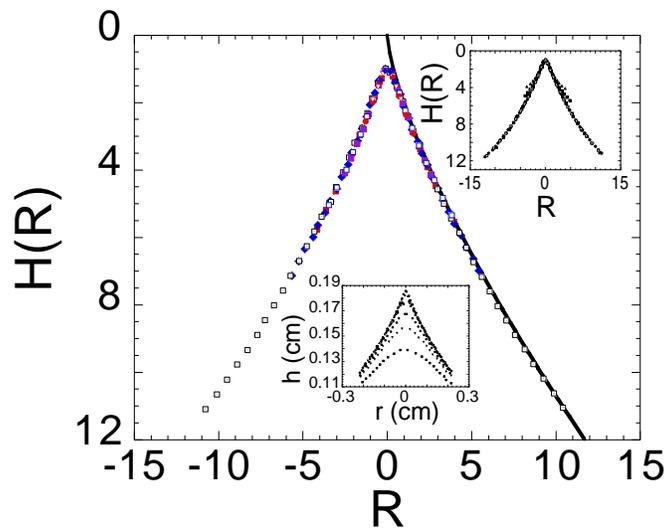}}}
\caption[]{
The scaled hump profiles.  The lower inset shows eight profiles taken 
from the $S = 0.830$ data set.  The main figure shows the same profiles after scaling.  The 
solid line corresponds to a power law of the form $R^{0.72}$.  In the upper inset we compare 
the universal curves for the $S = 0.830$ cm, $0.613$ cm, $0.508$ cm, $0.381$ cm, $0.255$ cm data 
sets.}
\label{fig:similarity}
\end{figure}

Here, we have shown that for the three-dimensional selective withdrawal system even when both 
fluids are viscous ($\eta > 1$ St for both fluids) the effects of a singularity manifest 
themselves in the scaling of the hump profiles.  Furthermore, preliminary experiments 
show that a reduction of the lower fluid viscosity to 0.01 St has little effect in determining 
the final curvature of the hump tip or equivalently, how close the system is to forming a 
cusp.  This suggests that for our 3-D problem, either the effects of the lower fluid 
viscosity enter as a higher-order perturbation to the profile shapes, or a different 
mechanism underlies the avoidance of the cusp formation.  If the latter scenario is correct, 
it may be possible for the system to manifest the singularity at a finite lower fluid 
viscosity.  In either case, determining which variables affect how close the system 
approaches the singularity would allow for control of the maximum hump curvature and 
minimum spout diameter.  This control could then be used to advance technologies such 
as coating microparticles \cite{Itai_coating}, creating mono-dispersed 
micro-spheres \cite{Ganan-Calvo98}, and emulsification through tip 
streaming \cite{Eggers01,Sherwood} which take advantage of the selective 
withdrawal geometry.

	We are grateful to W. W. Zhang, S. Venkataramani, J. Eggers, H. A. Stone, T. J. 
Singler, J. N. Israelachvili, C. C. Park, S. Chaieb, S. N. Coppersmith, T. A. Witten, L. P. 
Kadanoff, R. Rosner, P. Constantin, R. Scott, T. Dupont H. Diamant, and V. C. Prabhakar for 
sharing their insights.  This research 
is supported by the University of Chicago (MRSEC) NSF DMR-0089081 grant
%\bibliography{S_W_PRL}
%\bibliography{/home/icohen/icohen/Selective_Withdrawal/PRL_paper/S_W_PRL.bib} 

\begin{thebibliography}{10}
\bibitem{Svs.Q}
For systems with a different upper fluid viscosity, we find that the power-law
  dependence changes. There is a vast literature which focuses on explaining
  the Su vs. Q dependence (e.g. \cite{JRL89}). We therefore defer the detailed
  discussion of how currently available scaling predictions compare with the
  data to \cite{tobe}.
\bibitem{Surfac_disc}
Note that the transition from spout to hump is also discontinuous indicating
  that there is a curvature cut-off for the spout states as well. Since
  surfactants are continuously being removed from the interface when the system
  is in the spout state, the presence of surfactants in our system is not
  enough to account for the discontinuous nature of the transition or
  equivalently the values of the curvature cut-offs. More generally, we find
  that surfactant effects do not significantly influence the results presented.
  A more detailed discussion is presented in \cite{tobe}.
\end{thebibliography}
%calls the file containing references
%to add more files put a comma (,) and then the next file

\end{document}